\documentclass[aps,prl,twocolumn,showpacs,superscriptaddress,preprintnumbers,amsmath,amssymb,longbibliography]{revtex4-1}

\usepackage{dcolumn}
\usepackage{bm}
\usepackage[normalem]{ulem}    
\usepackage{xcolor}
\usepackage{gensymb}   
\usepackage{physics}
\usepackage{tikz}
\usepackage{mathtools}
\usepackage[normalem]{ulem}  
\usepackage[pdftex,colorlinks=true,pdfstartview=FitV,linkcolor=blue,citecolor=blue,urlcolor=blue]{hyperref}

\begin{document}
\title{Metallurgy at the nanoscale: domain walls in nanoalloys}
\author{Gr\'{e}goire Breyton}
\affiliation{Universit\'{e} Paris Cit\'{e}, Laboratoire Mat\'{e}riaux et Ph\'{e}nom\`{e}nes Quantiques (MPQ), CNRS-UMR7162, 75013 Paris, France}
\affiliation{Universit\'{e} Paris-Saclay, ONERA, CNRS, Laboratoire d'\'{e}tude des microstructures (LEM), 92322, Ch\^{a}tillon, France}
\author{Hakim Amara}
\affiliation{Universit\'{e} Paris-Saclay, ONERA, CNRS, Laboratoire d'\'{e}tude des microstructures (LEM), 92322, Ch\^{a}tillon, France}
\affiliation{Universit\'{e} Paris Cit\'{e}, Laboratoire Mat\'{e}riaux et Ph\'{e}nom\`{e}nes Quantiques (MPQ), CNRS-UMR7162, 75013 Paris, France}
\author{Jaysen Nelayah}
\affiliation{Universit\'{e} Paris Cit\'{e}, Laboratoire Mat\'{e}riaux et Ph\'{e}nom\`{e}nes Quantiques (MPQ), CNRS-UMR7162, 75013 Paris, France}
\author{Christine Mottet}
\affiliation{Aix-Marseille University/CNRS, CINaM UMR 7325, Campus de Luminy, Marseille 13288, France}
\author{Riccardo Gatti}
\affiliation{Universit\'{e} Paris-Saclay, ONERA, CNRS, Laboratoire d'\'{e}tude des microstructures (LEM), 92322, Ch\^{a}tillon, France}
\author{J\'{e}r\^{o}me Creuze}
\affiliation{Universit\'{e} Paris-Saclay, Institut de Chimie Mol\'{e}culaire et des Mat\'{e}riaux d'Orsay, UMR 8182, 91190, Orsay, France}
\author{Adrien Moncomble}
\affiliation{Universit\'{e} Paris Cit\'{e}, Laboratoire Mat\'{e}riaux et Ph\'{e}nom\`{e}nes Quantiques (MPQ), CNRS-UMR7162, 75013 Paris, France}
\author{Damien Alloyeau}
\affiliation{Universit\'{e} Paris Cit\'{e}, Laboratoire Mat\'{e}riaux et Ph\'{e}nom\`{e}nes Quantiques (MPQ), CNRS-UMR7162, 75013 Paris, France}
\author{Nathaly Ortiz Pe\~{n}a}
\affiliation{Universit\'{e} Paris Cit\'{e}, Laboratoire Mat\'{e}riaux et Ph\'{e}nom\`{e}nes Quantiques (MPQ), CNRS-UMR7162, 75013 Paris, France}
\author{Guillaume Wang}
\affiliation{Universit\'{e} Paris Cit\'{e}, Laboratoire Mat\'{e}riaux et Ph\'{e}nom\`{e}nes Quantiques (MPQ), CNRS-UMR7162, 75013 Paris, France}
\author{Christian Ricolleau}
\affiliation{Universit\'{e} Paris Cit\'{e}, Laboratoire Mat\'{e}riaux et Ph\'{e}nom\`{e}nes Quantiques (MPQ), CNRS-UMR7162, 75013 Paris, France}

\begin{abstract}
In binary alloys, domain walls play a central role not only on the phase transitions but also on their physical properties and were at the heart of the 70's metallurgy research. Whereas it can be predicted, with simple physics arguments, that such domain walls cannot exist at the nanometer scale due to the typical lengths of the statistical fluctuations of the order parameter, here we show, with both experimental and numerical approaches how orientational domain walls are formed in CuAu nanoparticles binary model systems. We demonstrate that the formation of domains in larger NPs is driven by elastic strain relaxation which is not needed in smaller NPs where surface effects dominate. Finally, we show how the multivariants NPs tend to form an isotropic material through a continuous model of elasticity.
\end{abstract}

\maketitle

Fundamental understanding of phase transformation in metallurgy is crucial for many industrial applications. In this context of materials science, this refers to a broad spectrum of physical phenomena strongly influenced by the presence of imperfections that all crystals contain (point, line, surface or volume defects). In the specific case of A$_{x}$B$_{1-x}$ bimetallic alloys research experienced an unprecedented boom in the 1960's, particularly with the study of first-and second-order phase transformations in relation to a singular type of defects present in solids, known as antiphase boundaries. These domain walls appear during the ordering process and are due to the presence of different sublattices which can be occupied either by A or B atom species during the nucleation stage, i.e. the translational symmetry is broken when going from one domain to another through this interface~\cite{Khachaturyan1983, Ducastelle1991, Reynaud1982}. In first order phase transitions, by increasing temperature close to the transition one, the domain walls are progressively transformed into two order/disorder interfaces separate by the disordered phase, so-called wetting phenomena. These new interfaces are associated with two characteristic lengths. The first one is the length $\ell$ of the disorder layer and the second one is the correlation length ($\xi$) defining the characteristic distance of order parameter fluctuations and their ratio diverges at the transition temperature~\cite{Leroux1990a, Leroux1990b, Ricolleau1992}. Except in the vicinity of the transition, these distances in bulk alloys remain very small (only a few interatomic spacings) compared to the size of domain walls explaining the stabilization of such ordered defects in materials.
At the nanometer scale, statistical fluctuations of the order parameter can be of the same order of magnitude as the small size of the considered nano-objects meaning that for nanoparticles (NPs) withstanding chemically-ordered structures, domain walls should therefore not exist. However, in the specific case of bulk anisotropic phases (L1$_{0}$, DO$_{22}$, DO$_{23}$, $\cdots$), the order-disorder phase transition is assisted by the formation of a multiple orientational domain wall microstructures with well-defined orientational relationships. It has been demonstrated that this organization of the ordered domains relaxes the bulk elastic strain energy, since the multidomain morphology approximates an invariant plane strain~\cite{Zhang1991}. Regarding NPs adopting similar anisotropic structure, the situation is not so obvious since the vicinity of free surfaces allows the fast relaxation of the elastic misfit, but it is not always the case. Indeed, it is known that non-crystalline structures such as icosahedral or multi-twinned NPs are stabilized by the presence of defects~\cite{Baletto2002, Nelli2023}. This raises the question of whether order defects exist in finite-sized alloys such as bimetallic NPs. Interestingly, despite extensive studies in the 1980's on those types of defects, particularly driven by the advancement of electron microscopy (in case of Co-Pt~\cite{Leroux1990b}, Cu-Au~\cite{Fisher1961, Hych1997}, Cu-Pd~\cite{Ricolleau1992}, Fe-Al~\cite{Swann1972,Lefloch1998}, $\cdots$), this fundamental metallurgy question has never been addressed at the nanoscale. 

\begin{figure*}[htbp!]
  \includegraphics[width=1.0\linewidth]{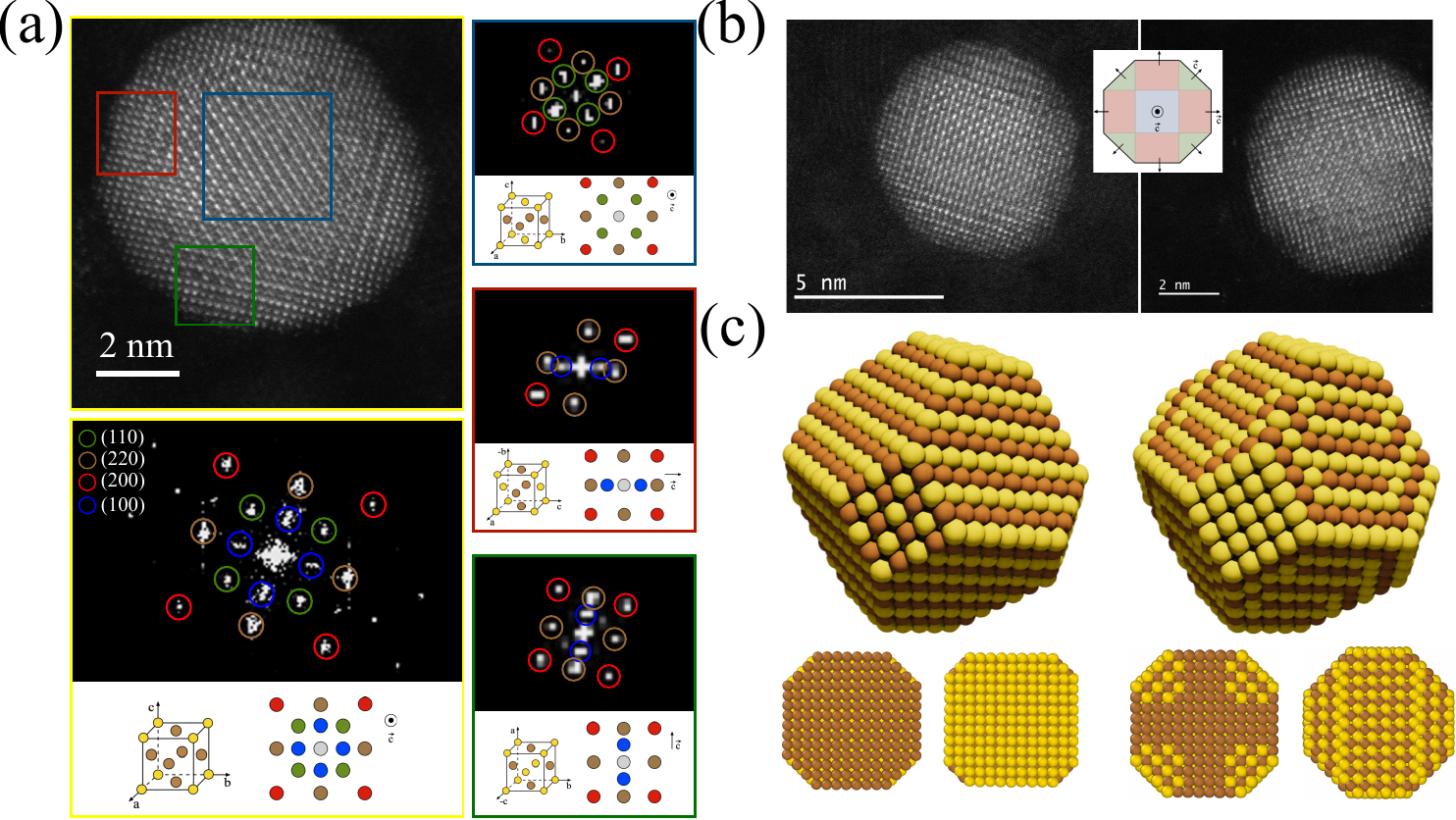}
  \caption{(a) HR-STEM image of a Cu$_{50}$Au$_{50}$ NP (mean size of 8 nm) oriented along the [001] axis (top) and its global diffraction pattern (bottom) corresponding to a L1$_{2}$ phase. The numerical diffraction pattern of the areas corresponding to the NP core (blue) and sides (red and green) are presented in the right panel. Note that in the diffraction scheme, the rectangles have been slightly oversized to illustrate the tetragonality. (b) Examples of Cu$_{50}$Au$_{50}$ NPs presenting multi-domain walls. (c) Schematic representations of the with mono-(left) and multi-variants (right) L1$_{0}$ NPs where some slice views are presented. Au atoms are in yellow and Cu atoms are in brown.}
    \label{fig:Figure_1}
\end{figure*}

Given the importance of orientational domain walls in wide range of phenomena such as phase transition or growth mechanisms at the nanoscale as well as the physical properties of NPs, we investigate such defects by combining high-resolution transmission  electron microscopy (HRTEM) experiments and atomic-scale simulations. We thus demonstrate the existence of orientation antiphase boundaries in case of large NPs (beyond 5 nm diameter) whose driving force is exactly the same as in bulk alloys, namely the minimization of elastic energy. Among all binary systems, CuAu is chosen as a model alloy since it is one of the most anisotropic material with a large tetragonalisation in the L1$_0$ ordered phase (low value of the $c/a$ factor) and where three types of ordered rotational domains can coexist corresponding to different tetragonal axis in various directions. To determine if there is a size effect, NPs ranging from 2 to 8 nm were studied. This range was chosen based in the known deviations from bulk properties such as the melting temperature~\cite{Buffat1976}, the surface energy~\cite{Amara2022}, or the order/disorder transition temperature~\cite{Alloyeau2009}.

Fig.~\ref{fig:Figure_1}a shows a High Angle Annular Dark Field (HAADF) image of a Cu$_{50}$Au$_{50}$ NP oriented along the [001] zone axis with a mean size of 8 nm acquired in scanning TEM (STEM) mode and its electron diffraction pattern obtained from numerical Fast Fourier Transform (FFT) analysis. Synthesis and TEM characterizations are given in Sec. I of the Supplementary Material. According to the extinction rules of the L1$_{0}$ phase, forbidden $hkl$ reflections occur whenever $h$ and $k$ have different parity for a given $l$. Thus, the only possible superstructure reflection along this zone axis, is the 110 ones. However, 100 and 010 reflections (and equivalents) are observed on the FFT of the whole NP. This exhibits a characteristic [001] zone axis diffraction of the L1$_{2}$ phase where all the superstructure reflections are present (Fig.~\ref{fig:Figure_1}a). This result is very surprising since the measured composition of the NP by Energy Dispersive X-ray Spectroscopy (EDX) gives a Cu$_{40}$Au$_{60}$ composition and not Cu$_{25}$Au$_{75}$ as expected in case of the L1$_{2}$ phase (see Sec. II of the Supplementary Material). To elucidate the physical origin of such observation, we conduct local structure analysis. The FFTs of the areas corresponding to the NP core and sides are shown in Fig.~\ref{fig:Figure_1}a. Due to the tetragonality of the L1$_{0}$ phase, three orientational domains can appear, one with the $\mathbf{c}$ axis perpendicular to the substrate and two others with the $\mathbf{c}$ axis in the substrate plane and perpendicular to each other. The diffraction of the core of the NP shows 110 superstructures reflections characteristic of the L1$_{0}$ phase oriented along the [001] axis (Fig.~\ref{fig:Figure_1}a). The sides of the NP exhibit diffraction pattern of the L1$_{0}$ structure viewed along the [100] and [010] directions with 001 superstructure reflections. These two domains have their $\mathbf{c}$ axis perpendicular to each other. This NP is then formed by ordered L1$_{0}$ orientational variants in coexistence, which is observed for the first time at the nanometer scale. This analysis explains the apparent conflicting results discussed beforehand. The L1$_{2}$-like diffraction from the whole NP corresponds in fact to the superposition of the diffraction of the three orientational domains of the L1$_{0}$ phase that coexist within the NP (see Sec. III of the Supplementary Material). As shown in Fig.~\ref{fig:Figure_1}b, the presence of multivariants is characterized by clearly identifiable planes with alternating contrasts oriented according to the three possible variants. Consequently, the very in-depth HRTEM analysis revealed a unique structure, which remains stable during annealing, that differs significantly from the "classical" L1$_{0}$ structure, shown in Fig.~\ref{fig:Figure_1}c, where schematic representations of the NP are presented. Besides, the slice views highlight strong differences within the NP, where the multivariant structure differs from the alternating Cu and Au planes typical of the monovariant one. \\
\begin{figure}[htbp!]
  \includegraphics[width=1.0\linewidth]{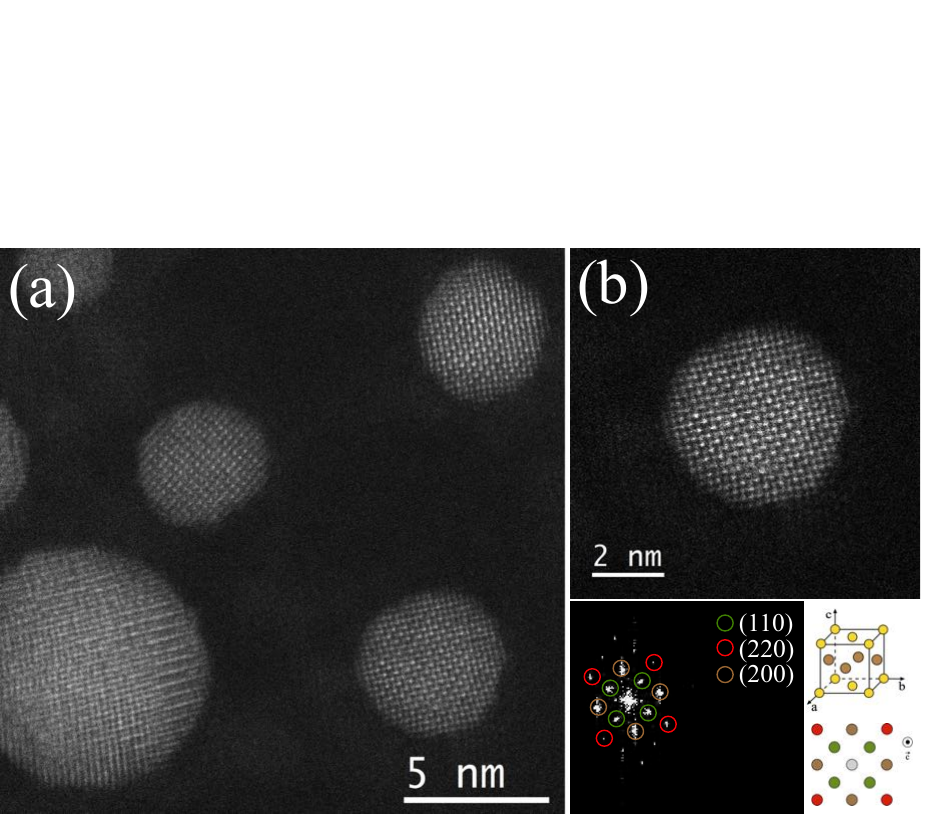}
  \caption{(a) HR-STEM image revealing the presence of a large NP with wall domains and smaller NPs of the monodomain type. (b) HR-STEM image of a Cu$_{50}$Au$_{50}$ NP (mean size of 3 nm) oriented along the [001] axis (top) and its FFT (bottom) corresponding to a L1$_{0}$ phase.}
    \label{fig:Figure_2}
\end{figure}

Strikingly, this coexistence is size dependent and only present in particles with mean diameter above about 4 nm. In contrast, smaller NPs appear as monodomain (Fig.~\ref{fig:Figure_2}). Furthermore, it is worth pointing out that this multidomain feature has never been observed in the case of NPs adopting an isotropic L1$_{2}$ phase. A typical example is presented in case of CuAu$_{3}$ composition where no contrast specific to multivariants is identified (see Sec. IV of the Supplementary Material). Accordingly, the size effects and the monovariant structure of the L1$_{2}$ phase highlighted experimentally strongly suggest that elasticity can be the driving force to stabilize mutlivariant domains. \\

In order to further understand the atomic mechanisms at the origin of the coexistence of multiple variants at the nanoscale, we perform Monte Carlo (MC) simulations in the canonical ensemble using a $N$-body potential derived from the second moment approximation (SMA) of the tight-binding (TB) scheme~\cite{Ducastelle1970, Rosato1989}. Note that the simulations are performed at low enough temperatures to ensure that the NPs are in a L1$_{0}$ ordered state as in experiments. Further details regarding the TB-SMA potential and the MC calculations are given in Sec. V of the Supplementary Material. 
\begin{figure}[htbp!]
  \includegraphics[width=1.0\linewidth]{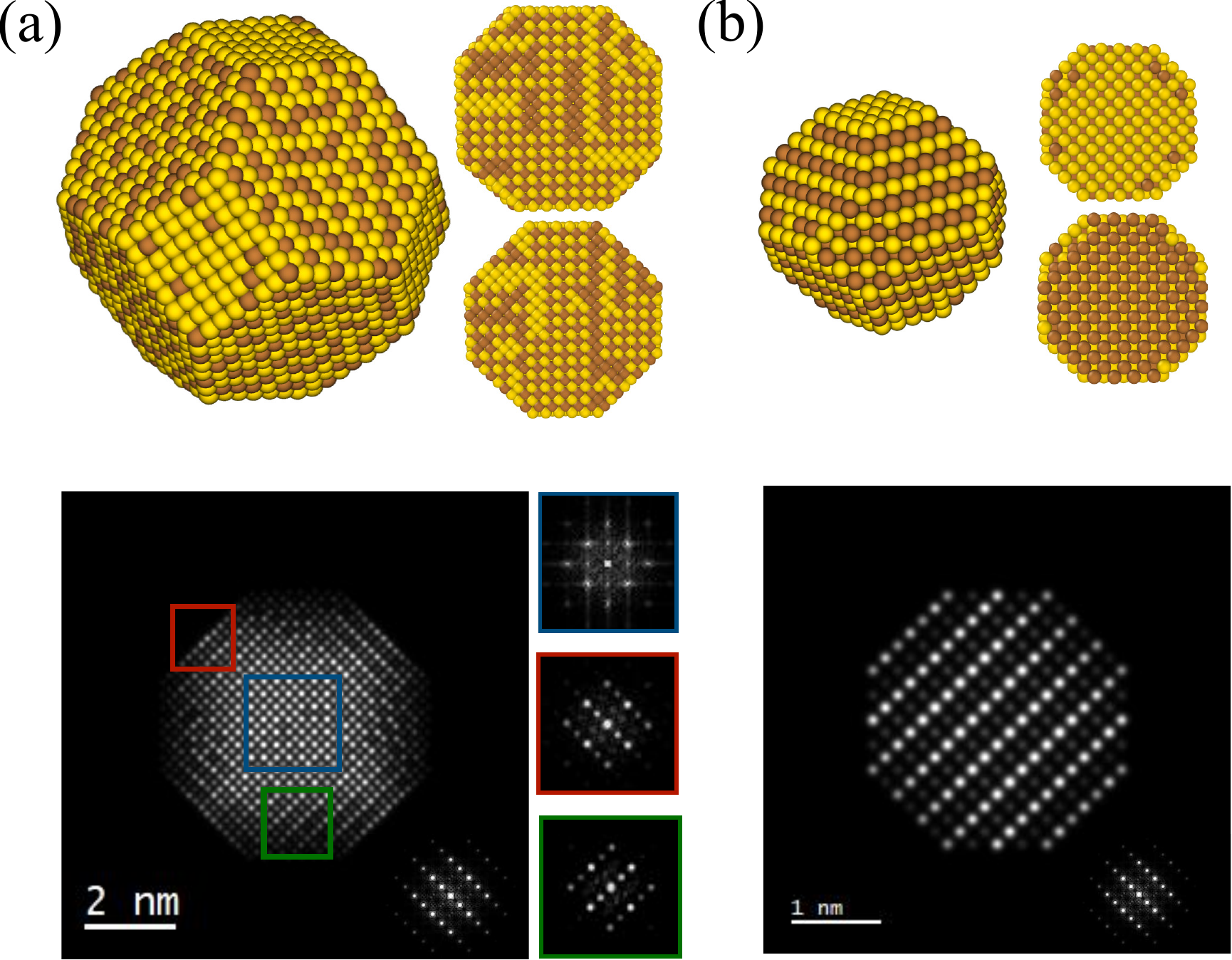}
  \caption{Top: Equilibrium configurations of Cu-Au NPs after performing MC simulations and their local diffraction patterns. Global and cross-section views of a characteristic equilibrium configuration are presented for a NP containing (a) 6266 atoms where multivariant domains are identified and (b) 1289 atoms where monovariant domain is observed. Au atoms are in yellow and Cu atoms are in brown. Bottom: HRSTEM simulations images of the corresponding configurations imaged along the [001] zone axis for the 6266 atoms NP and [100] zone axis for the 1289 atoms NP. Diffraction patterns are given in inset.}
    \label{fig:Figure_3}
\end{figure}
We considered truncated octahedron Cu$_{x}$Au$_{1-x}$ NPs containing 1289 and 6266 atoms. This corresponds to NP sizes around 3 and 6 nm respectively, close to the size range explored experimentally. We first focus on the largest NP with a composition similar to that studied experimentally, i.e. Cu$_{40}$Au$_{60}$. After a complete MC relaxation of the system, the final structure is strongly perturbed, especially the monovariant configuration is no longer present (see Fig.~\ref{fig:Figure_3}a). However, it is difficult to identify whether multi-variant domains coexist within the NP. The cross-section views seem to display an organization close to a multi-variant structure, but do not allow one to be entirely conclusive. To take the structural analysis a step further, HRSTEM images are simulated and quantitatively analysed by FFT, as done experimentally. Fig.~\ref{fig:Figure_3}a (bottom) shows the simulated HAADF high resolution STEM image of the same NP using the Dr Probe Software provided by Barthel~\cite{Barthel2018} (see technical details in Sec. VI of the Supplementary Materials). Structural features observed on this simulated nanoparticle are the same as the experimental ones. The whole NP exhibit a L1$_{2}$-like diffraction pattern. At the center of the NP, there is one L1$_{0}$ variant oriented with the $\mathbf{c}$ axis perpendicular to the plane of the figure while on the sides two variants with the $\mathbf{c}$ axis in the fictitious substrate plane and perpendicular to each other are present. Furthermore, a smaller particle of 3 nm diameter is considered. In this particular case, the L1$_{0}$-type structure remains stable after MC relaxation, again carried out after heating then cooling, above and below, the order-disorder phase transition (see Fig.~\ref{fig:Figure_3}b). The slice views clearly show the alternate Cu and Au planes along the $c$-axis typical of a L1$_{0}$ structure. Thus, the simulated image corresponding to the NP of 3 nm diameter is clearly the one of a monodomain L1$_{0}$ NP.\\

\begin{figure}[htbp!]
  \includegraphics[width=1.0\linewidth]{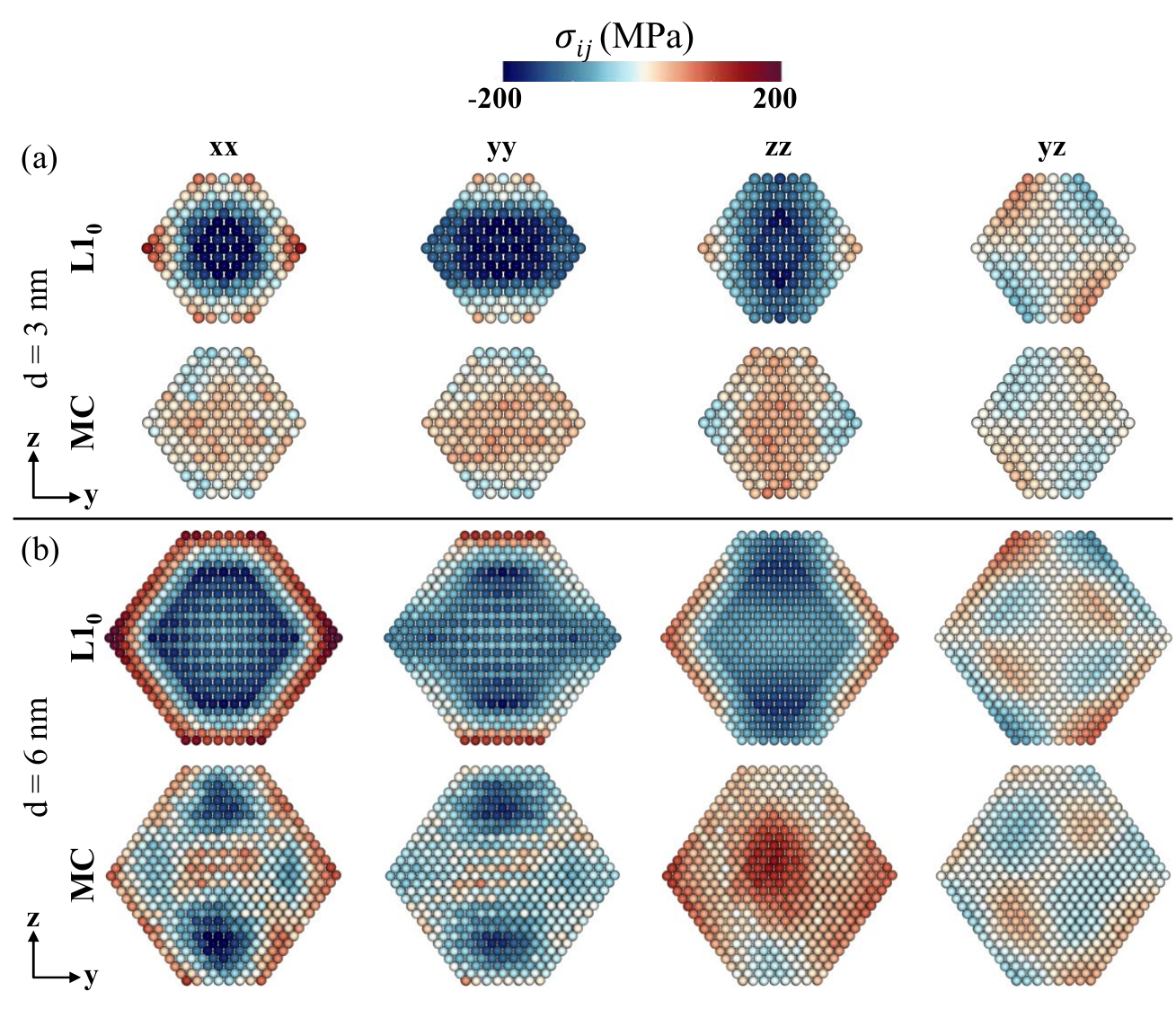}
  \caption{Internal stress distribution in Cu-Au NPs of different sizes. $\sigma_{xx}$, $\sigma_{yy}$, $\sigma_{zz}$ and $\sigma_{yz}$ components are plotted for 3 nm diameter NPs in panel (a) and for 6 nm diameter NPs in panel (b) for L1$_{0}$ structure and after MC relaxation.}
    \label{fig:Figure_4}
\end{figure}

To capture the impact of elastic relaxation on the development of multivariant domains, we analyse the local stress ($\sigma_{ij}$) of each atom using the open source large-scale atomic/molecular massively parallel simulators (LAMMPS) package~\cite{Plimpton1995} by means of the stress maps. Only four stress components are discussed ($\sigma_{xx}$, $\sigma_{yy}$, $\sigma_{zz}$ and $\sigma_{yz}$) since  $\sigma_{xy}$ and $\sigma_{xz}$ are small. To evaluate size effect, the stress distributions have been calculated in case of small (3 nm) and large (6 nm) NPs and are displayed in Fig.~\ref{fig:Figure_4} where negative and positive values correspond to compressive and tensile stress, respectively. The internal stress maps shows large stress values in the L1$_{0}$ structures due to the mismatch between Au and Cu layers. The analysis of the NPs after MC relaxation reveals that for both sizes a decrease of internal stress values due to the reorganisation of Au and Cu distribution inside the NPs as show in Fig.~\ref{fig:Figure_4} (a) and (b). In particular, it is clear that the organization of ordered domains revealed in the largest NP stems from the redistribution of the internal stress and the relaxation of the elastic strain energy. Indeed, the maps corresponding to the diagonal terms are more homogeneously distributed within the NP compared to the case of the monovariant one. Regarding, the non-diagonal terms, we can clearly see that the presence of multi-domains significantly reduces their contribution displaying values close to zero. To conclude, detailed analysis of the elastic distribution within the NP in form of stress map as well as histogram (see Sec. VII. of the Supplementary Materials) shows that the presence of the variants isotropises the stress contribution by adopting a homogeneous distribution coupled with specific components being null.\\

Interestingly, the formation of multivariant domains is not limited to the CuAu system. CoPt, that also shows a L1$_{0}$ order/disorder phase transition, exhibits the same behavior (see Sec. VIII of the Supplementary Materials). Since this mechanism is generalized, we here proposed a model based on elastic theory to explain the driving forces to stabilize the multidomain NPs. 
\begin{figure}[htbp!]
  \includegraphics[width=1.0\linewidth]{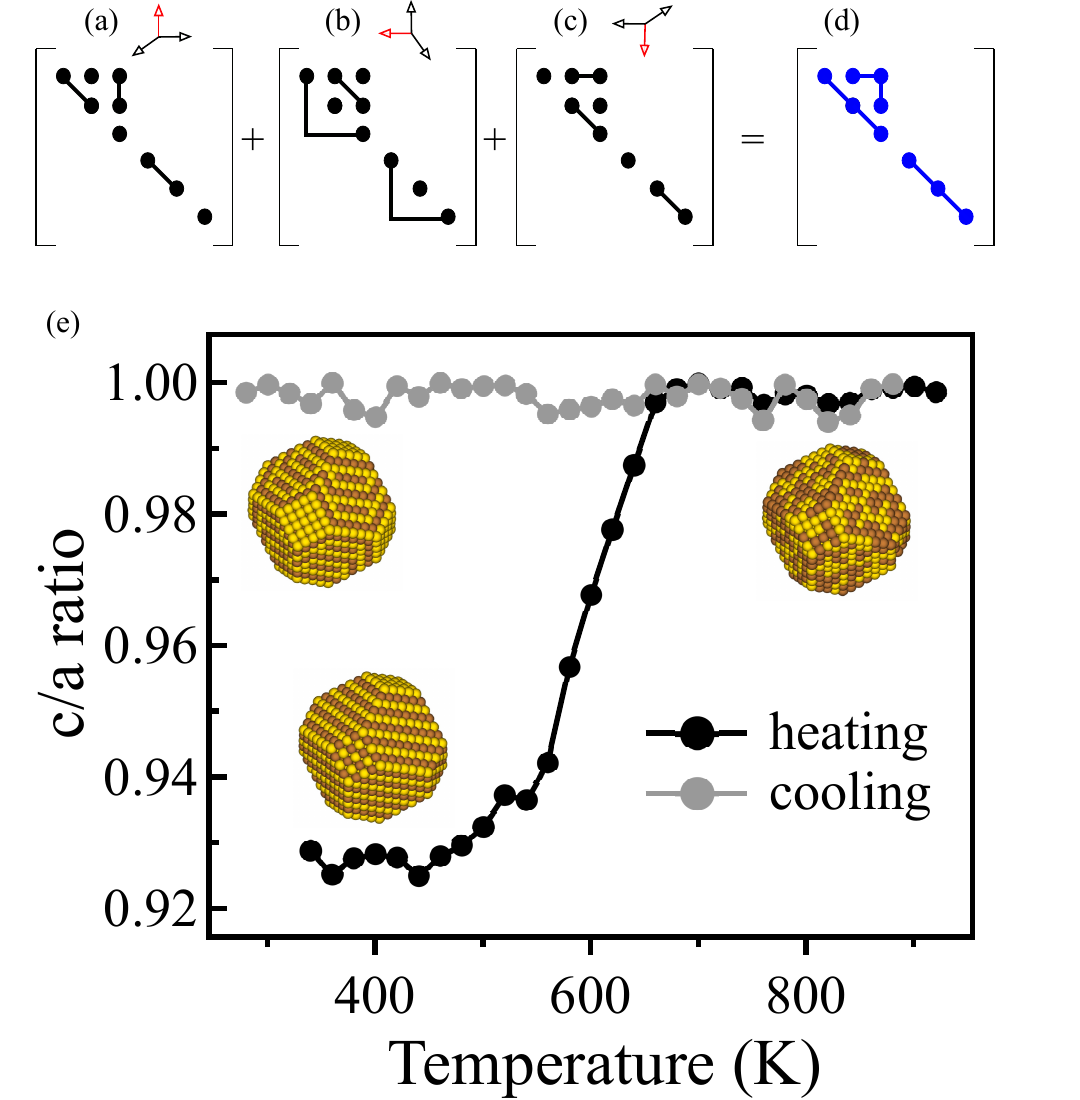}
  \caption{The patterns of the elastic stiffness tensor for (a), (b) and (c) corresponding to the L1$_{0}$ structures with different orientations (the $\mathbf{c}$ axis is in red) and (d) the cubic system. Only the non-zero components are indicated and equal components are connected. (e) Evolution of the $c/a$ ratio of the CuAu NP during MC simulations under heating and cooling.}
    \label{fig:Figure_6}
\end{figure}
Hence, we first consider the elastic stiffness tensor of the L1$_{0}$ structure in an orthonormal basis ($Oxyz$) where the 4-fold symmetry axis is chosen parallel to $Oz$ (see Fig.~\ref{fig:Figure_6}a). Let us evaluate now this elasticity tensor in a new orthonormal basis obtained by a rotation by $\pi/2$ around $Ox$ direction. This operation transforms the components $x_{1}$, $x_{2}$, $x_{3}$ of a vector into components $x_{1}^{'}=x_{1}$, $x_{2}^{'}=x_{3}$, $x_{3}^{'}=-x_{2}$. In this new axis coordinates, according to the rule of transformation of the tensor coefficients during a change of orthonormal basis, the matrix representing the stiffness tensor in the Voigt's notation is modified as presented in Fig.~\ref{fig:Figure_6}b. Similarly, in a new basis obtained by $\pi/2$ around $Oy$ direction, the elasticity tensor has the form shown in Fig.~\ref{fig:Figure_6}c. If we consider now a material formed by an equal volume fraction of the three L1$_{0}$ variants with their $\mathbf{c}$ axis perpendicular two by two, the mean elastic tensor of the material will take the form of a cubic material (see Fig.~\ref{fig:Figure_6}d), i.e. a more isotropic system. This is in agreement with numerical analysis showing that in case of multivariant domains, local stresses are zero or homogeneous. As a result, this macroscopic model based on the theory of elasticity (see details in Sec. IX. of the Supplemental materials), reveals that this mechanism can be generalized to any bimetallic systems with a low temperature L1$_{0}$ structure.

\begin{table} [htbp!]
\begin{center}
\begin{tabular}{ |c |c| c |}
 \hline
 System & $c/a$ ratio & A$_{U}$ \\
 \hline
 \hline 
 CuAu & 0.926 & 2.88 \\  
 FePd & 0.966 & 0.77 \\
 FePt & 0.968 & 0.75 \\
 CoPt & 0.973 & 0.57 \\
\hline
\end{tabular}
\caption{\textbf{Descriptors of the anisotropy of the L1$_{0}$ tetragonal structure.} Values of the $c/a$ ratios and the generalized Zener coefficient A$_{u}$ and for the L1$_{0}$-type materials: CuAu, FePd, FePt and CoPt~\cite{Singh1993, Jain2013}.}
\label{tab:Table_1}
\end{center}
\end{table}

Regarding the literature, the formation of multivariants NPs was already observed in other L1$_{0}$ type systems like FePt~\cite{Bian1999, Bian2000, Sato2002, Li2015}, FePd~\cite{Sato2003, Sato2002} and CoPt~\cite{Sato2017, Chi2015}. However, the physical origin of the existence of such multivariants NP is not explained, or even not commented~\cite{Li2015}. By comparing our results with the configurations shown in all these works, it appears that depending on the $c/a$ ratio values of the L1$_{0}$ phase (Table~\ref{tab:Table_1}), and in agreement with the macroscopic demonstration given above, the extension of variants is more or less important. Indeed, for the CuAu system, the influence of the multivariants arrangement is clearly seen on the evolution of the mean $c/a$ ratio calculated over the NP (Fig.~\ref{fig:Figure_6}e). Starting from a monovariant NP and upon heating, the $c/a$ varies from 0.93 to 1 and allow to define the critical order/disorder transition temperature for a NP of 6 nm diameter. Upon cooling, the three variants nucleate and the mean $c/a$ ratio in the whole NP is closed to 1, i.e. equivalent to a cubic structure. This result validates the simple demonstration given above using the equivalent stiffness tensor of a multivariants NP, together with the lowering of the total elastic energy obtained from the atomistic simulation when transiting from monovariant to multivariants NP. As a result, the degree of anisotropy of the L1$_{0}$ structure with respect to a cubic structure can be described by the $c/a$ ratio of the tetragonal phase. Among the four systems, CuAu with a $c/a$ of 0.926 is the most anisotropic system and the volume extension of the variants is the most important in order to minimize the elastic energy and to tend to the one of an equivalent cubic system. Another more direct way to characterize the anisotropy of a material is looking at the stiffness components. Thus, Ranganathan \textit{et al.} introduced an universal anisotropy index, A$_{U}$, which takes into account all the stiffness components of the elastic stiffness tensor instead of the ratio of individual stiffness coefficients to define anisotropy~\cite{Ranganathan2008}. For the four alloys discussed here, the values of A$_{U}$, given in Table~\ref{tab:Table_1}, reflects perfectly the behavior of the $c/a$ ratio justifyying its use as a good descriptor of the anisotropy of the L1$_{0}$ tetragonal structure. \\

Bimetallic nanoparticles are fascinating materials whose unique properties are closely linked to their chemical order, i.e. core-shell, random, phase separation, Janus, ...~\cite{Ferrando2008, Ferrando2018} Here, we have observed, demonstrated and explained the existence of a new mixing pattern corresponding to the presence of rotational antiphase boundaries persisting beyond a certain size ($>$ 5 nm). Surprisingly, these ordered defects have never been discussed in the literature although HRTEM images have shown their presence. In addition to making an important contribution to the understanding of self-organization within a bimetallic NP, it is clear that such chemical ordering  will have fairly unique impact on various properties that will need to be studied or even taken into consideration to revisit and reinterpret previous data. Typical example is the magnetocrystalline anisotropy characteristic of equiatomic binary alloys known to be due to an L1$_{0}$ order that alternates pure planes along one direction. Yet studies have often discussed the effect of size or the influence of annealing on the magnetic properties of L1$_{0}$-type NPs~\cite{Wiedwald2010, Tournus2010, Alsaad2015,Dupuis2013, Nam2023}, and the relationship with the presence or not of domain walls deserves to be considered. Furthermore, the present contribution on the metallurgy at the nanoscale has also raised the question of whether order defects such as domain walls exist in finite-sized bimetallic alloys. Despite extensive studies in the 1980s on those kind of defects, this fundamental question has never been addressed in NPs. This work then constitutes a major step forward in the NPs community, emphasizing once again that these nano-objects are a truly exceptional playground for tackling and highlighting unexpected physical singularities. 

\section*{Supplemental material : \\Metallurgy at the nanoscale: domain walls in nanoalloys}

\textbf{Sec. I. Experimental details: synthesis and TEM analysis}\\

Cu$_{50}$Au$_{50}$ and Co$_{50}$Pt$_{50}$ NPs were prepared by alternated pulsed laser deposition (PLD) technique using a KrF excimer laser source in a high vacuum chamber under a pressure of $10^{-8}$ Torr~\cite{Alloyeau2007}. The NPs were grown by epitaxy on a NaCl substrate and deposited on a TEM carbon grid by the carbon replica technique~\cite{PierronBohnes2014, Breyton2023}. The nominal thickness is 0.8 nm, the growth temperature is fixed at a temperature below the order/disorder phase transition in order to obtain NPs in the ordered phase (L1$_{0}$ phase), and the laser frequency is 5 Hz. The NPs were imaged by using a double aberration corrected electron microscope (JEOL ARM 200F cold FEG) in STEM mode using the High Angle Annular Dark Field (HAADF) technique. Chemical analysis of the NPs composition was performed by Energy Dispersive X-ray Spectroscopy (EDX) at the single particle level. \\

\textbf{Sec. II. EDX spectra of CuAu NPs}\\

\begin{figure}[htbp!]
\includegraphics[width=1.0\linewidth]{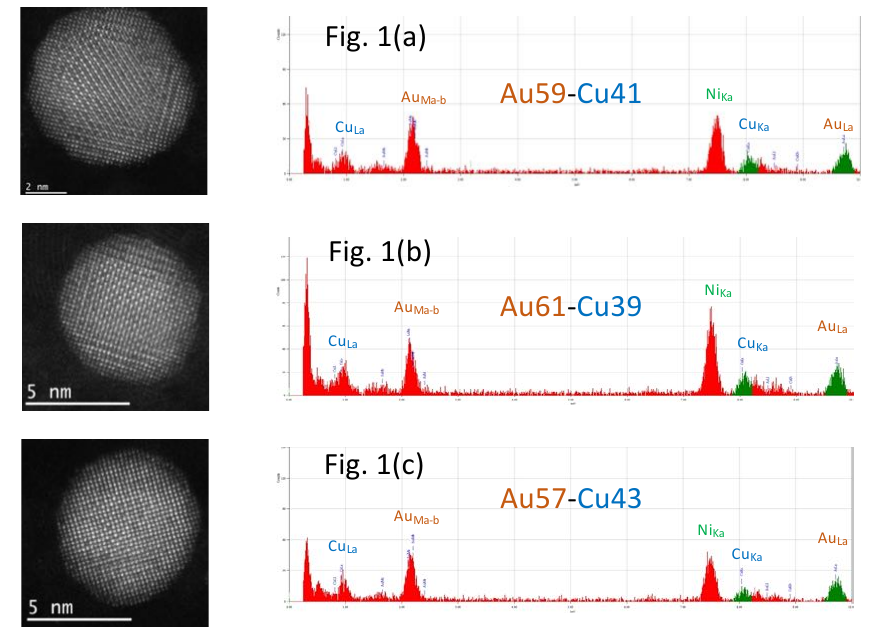}
  \caption{EDX spectra acquired on the different NPs presented in the Figure 1 of the manuscript.}
    \label{fig:Figure_S0}
\end{figure}
In Figure~\ref{fig:Figure_S0} are presented the EDX spectra measured on the NPs presented in Figure 1 of the manuscript. In all cases, the composition of the NPs by EDX gives a 60/40 composition in Cu and Au respectively. \\

\textbf{Sec. III. Diffraction patterns of L1$_{2}$ and the L1$_{0}$ phases along different directions }\\

\begin{figure}[htbp]
\includegraphics[width=1.0\linewidth]{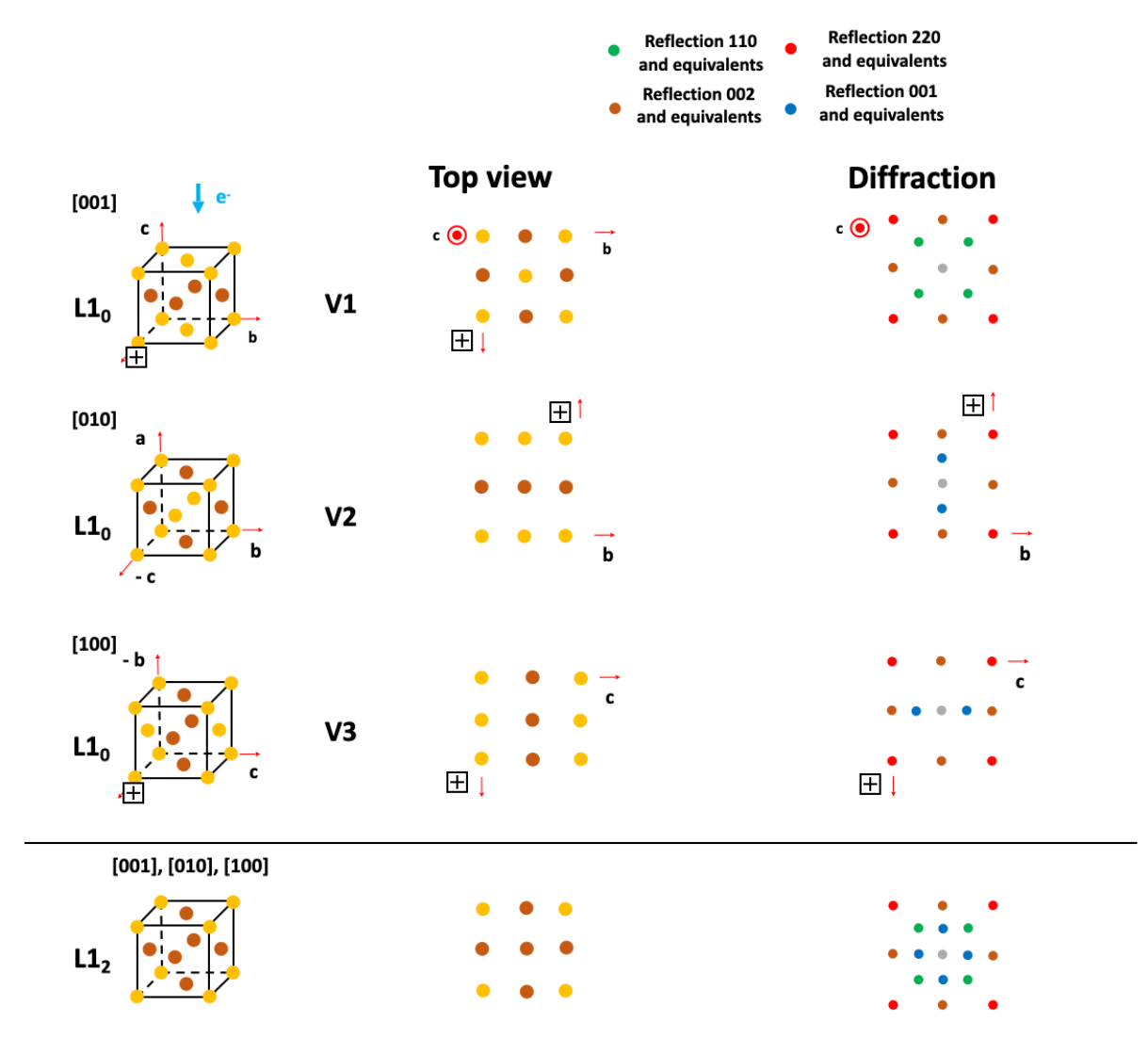}
  \caption{Drawings showing the different configurations for the L1$_{0}$ phase according to $\left[001\right]$, $\left[010\right]$ and $\left[100\right]$ directions. A top view and the associated diffractions are also shown. Higher-order structural reflections are not shown.}
    \label{fig:Figure_S1}
\end{figure}

The analysis presented in Figure~\ref{fig:Figure_S1} allows us to solve the enigma raised by the global diffraction of the particle shown in Figure 1 of the main manuscript. By superimposing the three diffraction patterns obtained for each L1$_{0}$ variant (which is equivalent to a global diffraction of the NP), we end up with reflections in all directions, making this diffraction identical to that of a L1$_{2}$ structure oriented according to $\left[100\right]$.\\

\newpage

\textbf{Sec. IV. Structural and chemical analysis of CuAu$_{3}$ NP in a L1$_{2}$ ordered phase}\\

\begin{figure}[htbp]
  \begin{center}
\includegraphics[width=1.0\linewidth]{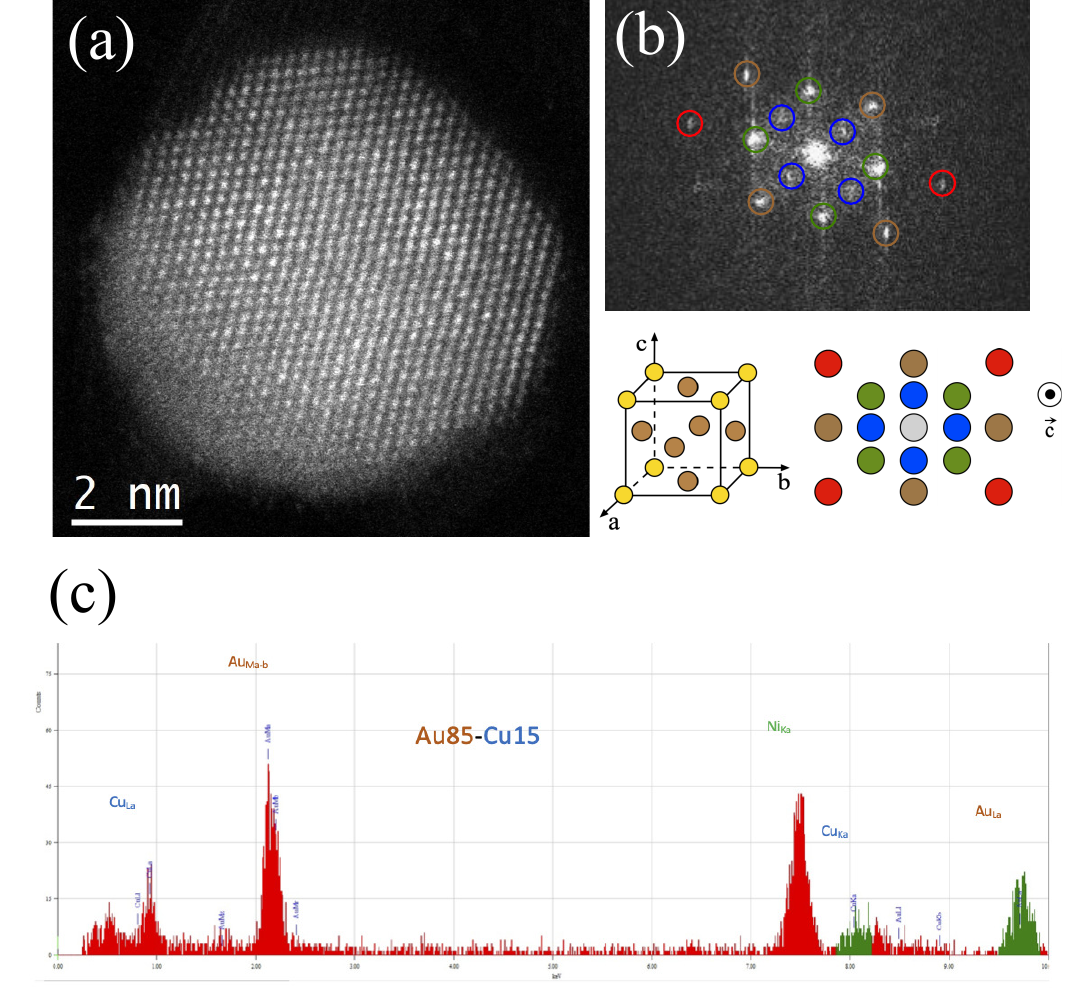}
  \caption{(a) HR-STEM image of a CuAu$_{3}$ NP (mean size of 6 nm) oriented along the [001] axis, (b) its global diffraction pattern corresponding to a L1$_{2}$ phase and (c) EDX spectra revealing the exact composition, Cu$_{15}$Au$_{85}$.}
    \label{fig:Figure_SM_L12}
   \end{center}    
\end{figure}

In Figure~\ref{fig:Figure_SM_L12} is presented the analysis of typical example of a Au$_{3}$Cu NP.

\textbf{Sec. V. MC simulations based on a TB interatomic potential}\\

The interatomic potential is included in a MC code in the canonical and semi-grand canonical ensembles to relax the structures at finite temperatures. The complete SMA model and parameter fitting procedure for pure metals and mixed interactions are detailed in Ref.~\cite{Breyton2023} for Cu-Au and in Ref.~\cite{Front2021} for Co-Pt. 

MC simulations are based on the Metropolis algorithm carried out in the isothermal-isobaric ensemble (number of atoms of each species, temperature and pressure are conserved - zero pressure for NM in vacuum !) also called canonical ensemble. To types of MC trials are applied: random atomic displacements and random atomic exchange between two atoms of different species at different sites. Each MC trial is accepted with the probability equal to 1 if the total energy of the system is lower after the trial, if not, the trial is accepted with a probability equal to $exp(-\frac{\Delta E}{k_BT})$ where $\Delta E$ is the energy difference of the system after the trial, $k_B$ the Boltzmann constant and $T$ the temperature. For a NP of $N$ atoms we typically run 2000 macrosteps after 800 of equilibration then the last 1200 macrosteps are used to calculate the ensemble averages. For each macrostep we apply $N$ chemical exchanges for $20\times N$ atomic displacements, randomly chosen respectively at each macrostep, which leads to about 30 millions to 120 millions of microsteps, depending on the size of the system (1289 or 6266 atoms). Heating and cooling starting from L1$_0$ ordered or disordered NP with a increment of temperature of 20 K, starting with the configuration obtained at the precedent temperature. Up to five different runs are performed in order to improve the statistic. This procedure ensures to get well converged equilibrium structures at low temperatures below the order/disorder temperature to be compared with experiment. 

The analysed structures were obtained after a complete relaxation procedure by increasing the temperature above the critical order/disorder temperature up to a complete chemical disordering of the NP, and slowly decreasing the temperature down to T = 150 K.\\

\textbf{Sec. VI. Parameter for the HRSTEM images }\\

Simulated HRSTEM images of the nanoparticles were generated using the Dr. Probe software~\cite{Barthel2018} based on the multislice algorithm. In order to have the simulated image as close as possible to the experimental ones, we need to set the parameters according to the characteristics of the microscope. In Table~\ref{tab:param-table} are presented the mean values of the aberration coefficients corresponding to the objective lens of the JEOL JEM-ARM200F spherical-aberration- corrected electron microscope equipped with a cold field emission gun operated at 200 kV used for the acquisition of the experimental images.
The semi-angle of convergence and annular detector radii were given by JEOL. The electron probe radius was taken to be equal to the STEM resolution of the microscope with aberration correctors. The values of the aberration were taken from the measurement done by the STEM aberration corrector software.

\begin{table}[htb]
    \centering
    \begin{tabular}{|c|c|}
        \hline
        Simulations Parameters & Value \\
        \hline
         Microscope energy & 200 kV \\
         Semi-angle of convergence & 30 mrad \\
         Electron probe radius & 0.4 \r{A} \\
         Annular detector inner radius & 68 mrad \\
         Annular detector outer radius & 280 mrad \\
         Defocus (C1) & -1.2 nm \\
         Spherical aberration (C3) & -1.95 $\mu$m \\
         \hline
    \end{tabular}
    \caption{The parameters correspond to a JEOL ARM 200F cold FEG  using a double aberration corrected electron microscope.}
    \label{tab:param-table}
\end{table}

\textbf{Sec. VII. Calculations of the local stress }\\
\begin{figure}[htbp!]
  \begin{center}
\includegraphics[width=0.70\linewidth]{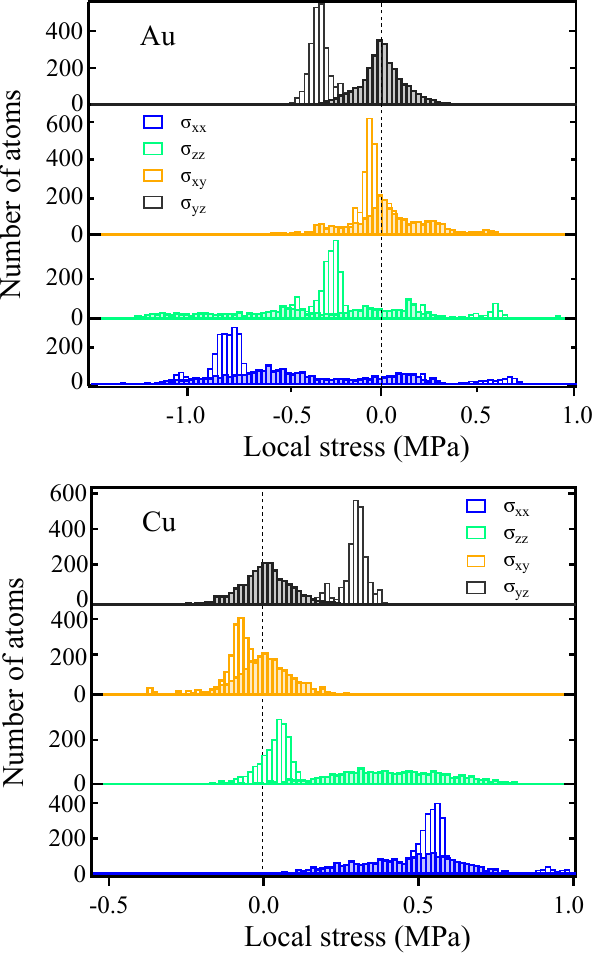}
\caption{Stress tensor in a histogram form (Cu and Au atoms) corresponding to the mono (empty) and multi-domain (filled) structures for the large NP.}
\label{fig:Figure_Stress}
   \end{center}
\end{figure}

Figure~\ref{fig:Figure_Stress} depicts the four components of the stress tensor in a histogram form (Cu and Au atoms) corresponding to the mono and multi-domain structures for the 6 nm diameter NP. From this analysis, a number of conclusions can be drawn. First, Au and Cu atoms behave differently, inducing mainly compressive and tensile stresses, respectively. This is simply due to the difference in atomic radius, which is around 13\% greater for Au. For both elements, the monovoriant structure exhibits fairly sharp distributions for all components, typical of a structure under stress, with especially very high stress values of $\sigma_{xx}$ around -0.80 MPa and 0.55 MPa for Au and Cu, respectively. For the multivariant structure, the analyzed landscape changes drastically. Hence, the distributions for the diagonal terms are no longer sharply defined but homogeneously distributed over a wide stress range of about 1 MPa. For the non-diagonal terms, we can clearly see that the presence of multi-domains significantly reduces their contribution with a distribution center of gravity close to zero. This is particularly well illustrated in the case of $\sigma_{yz}$ which drops from 0.30 MPa (absolute value) to almost zero for both Cu and Au.\\

\textbf{Sec. VIII. Multivariants in CoPt NPs }\\

\begin{figure}[htbp!]
\includegraphics[width=1.0\linewidth]{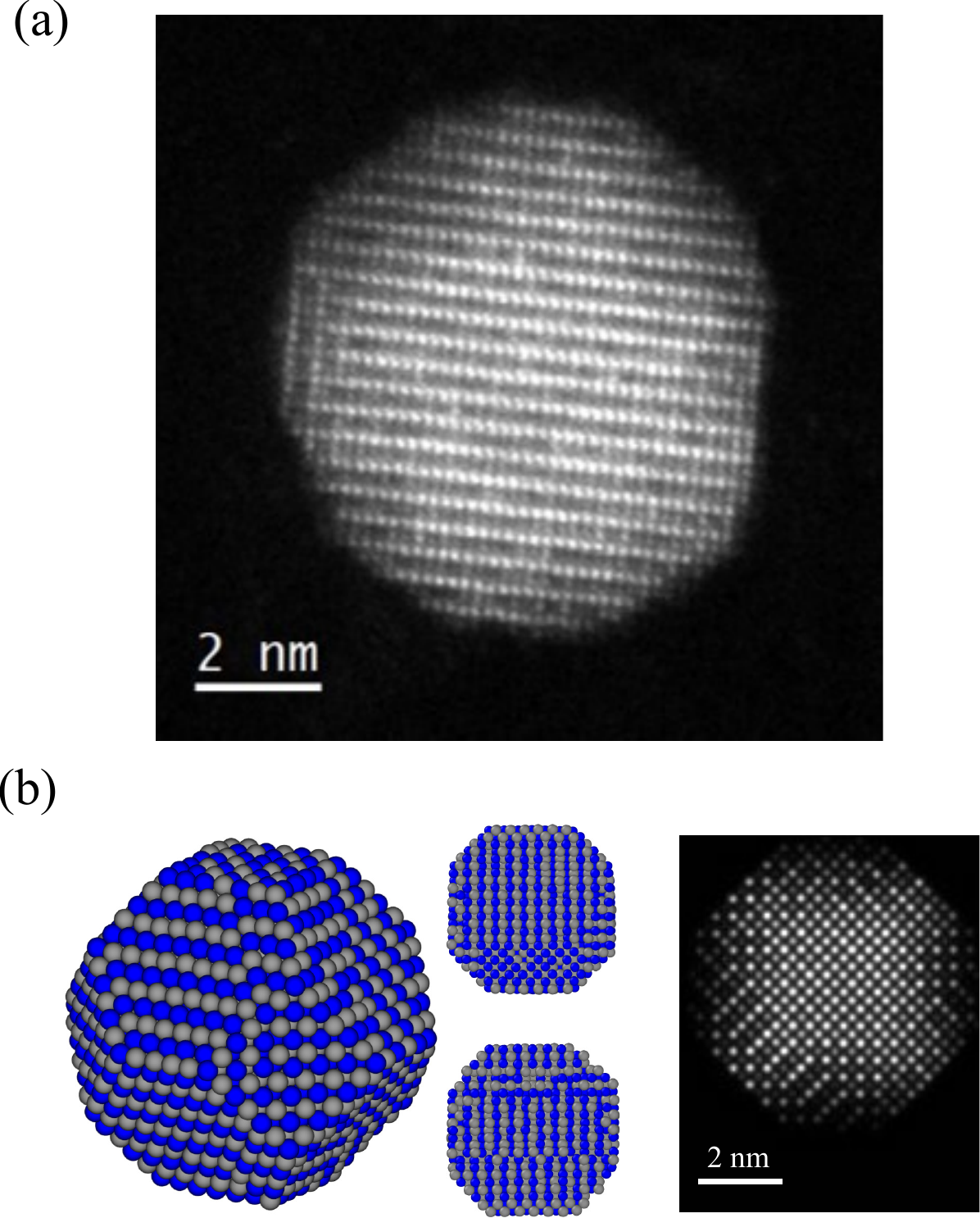}
\caption{(a) HRSTEM image of a CoPt NP exhibiting multivariant domains. (b) Equilibrium configurations of CoPt NPs after performing MC simulations and its calculated HRSTEM image. Global and cross-section views of a characteristic equilibrium configuration  are  presented  for a NP containing around 3500 atoms. Co atoms are in blue and Pt atoms are in grey.}
\label{fig:Figure_S2}
\end{figure}

In Figure~\ref{fig:Figure_S2}a. Co$_{50}$Pt50$_{50}$ NPs microstructure exhibits the same feature as in CuAu with the coexistence of two orientational domains. The only difference with CuAu is the lateral expansion of the variant with $\mathbf{c}$ axis parallel to the substrate plane which is smaller. This is also confirmed by MC simulations coupled to HRSTEM images calculations. As seen in Figure~\ref{fig:Figure_S2}b, the overall appearance of the NP does not reveal the mulivariant structure, nor do cross-sectional views. Here again, the HRSTEM calculation is crucial, enabling the unambiguous demonstration of the coexistence of different orientational domains.\\
\newpage 

\textbf{Sec. IX. Elasticity theory at the macroscopic scale }\\

By comparison to the stiffness tensor of the cubic system which needs only three elastic components ($C_{11}=C_{22}=C_{33}$, $C_{12}=C_{13}=C_{23}$ and $C_{44}=C_{55}=C_{66}$) to describe its mechanical properties, the one of the tetragonal L1$_{0}$ structure has 3 more $C_{\alpha\beta}$ components, namely $C_{33}$, $C_{13}$ and $C_{66}$ in an orthonormal basis ($Oxyz$) where the 4-fold symmetry axis is chosen parallel to $Oz$ leading to the following matrix where lines connecting some dots indicating that the corresponding coefficients are equal:

\begin{figure}[htbp!]
  \begin{center}
  \includegraphics[width=0.3\linewidth]{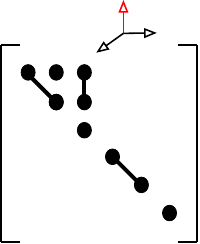}
  \end{center}
\end{figure}

Let us evaluate now this elasticity tensor in a new orthonormal basis obtained by a rotation by $\pi/2$ around $Ox$ direction. This operation, performed on the axis system transforms the components $x_{1}$, $x_{2}$, $x_{3}$ of a vector into components $x_{1}^{'}$, $x_{2}^{'}$, $x_{3}^{'}$ such that:

\begin{equation}
  \left\{
      \begin{aligned}
        x_{1}^{'}&=x_{1} \\
        x_{2}^{'}&=x_{3}\\
        x_{3}^{'}&=-x_{2}\\
      \end{aligned}
    \right.
\end{equation}

In this new axis coordinates, according to the rule of transformation of the tensor coefficients during a change of orthonormal basis, the new components of the stiffness tensor are written as:

\begin{equation}
  \left\{
      \begin{aligned}
        C_{11}^{'}&=C_{11} & C_{22}^{'}&=C_{33} & C_{33}^{'}&=C_{22}=C_{11}\\
        C_{12}^{'}&=C_{13}=C_{23} & C_{13}^{'}&=C_{12} & C_{23}^{'}&=C_{23}\\
        C_{44}^{'}&=C_{44}=C_{55} & C_{55}^{'}&=C_{66} & C_{66}^{'}&=C_{55} \\
      \end{aligned}
    \right.
\end{equation}
In such a case, the elasticity tensor in this basis has a new form:

\begin{figure}[htbp!]
  \begin{center}
  \includegraphics[width=0.3\linewidth]{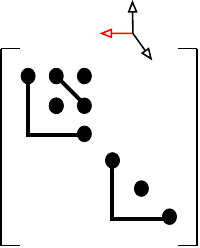}
  \end{center}
\end{figure}

Similarly, in a new basis obtained by $\pi/2$ around $Oy$ direction, the new components are written as:

\begin{equation}
  \left\{
      \begin{aligned}
        x_{1}^{'}&=-x_{3}\\
        x_{2}^{'}&=x_{2}\\
        x_{3}^{'}&=-x_{1}\\
      \end{aligned}
    \right.
\end{equation}
leading to the new components : 
\begin{equation}
  \left\{
      \begin{aligned}
        C_{11}^{'}&=C_{33} & C_{22}^{'}&=C_{22}=C_{11}\ & C_{33}^{'}&=C_{11}\\
        C_{12}^{'}&=C_{23}=C_{13} & C_{13}^{'}&=C_{13} & C_{23}^{'}&=C_{12}\\
        C_{44}^{'}&=C_{66} & C_{55}^{'}&=C_{55} & C_{66}^{'}&=C_{44}=C_{55} \\
      \end{aligned}
    \right.
\end{equation}
As a result, we finally obtain the following matrix : 
\bigskip
\begin{figure}[htbp!]
  \begin{center}
  \includegraphics[width=0.3\linewidth]{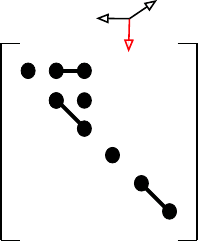}
  \end{center}
\end{figure}

If we consider now a material formed by an equal volume fraction of the three L1$_{0}$ variants with their $\mathbf{z}$ axis perpendicular two by two, the mean elastic tensor of the material will take the form:

\begin{equation}
  \left\{
      \begin{aligned}
        C_{11}^{*}&=&C_{11} + C_{11} + C_{33}&=& 2C_{11}+C_{33}\\
        C_{22}^{*}&=&C_{11} + C_{33} + C_{11}&=& 2C_{11}+C_{33} \\
        C_{33}^{*}&=&C_{33} + C_{11} + C_{11}&=& 2C_{11}+C_{33}\\
        \cdots\\
      \end{aligned}
    \right.
\end{equation}
This results in the corresponding matrix form:
\begin{figure}[htbp!]
  \begin{center}
  \includegraphics[width=0.3\linewidth]{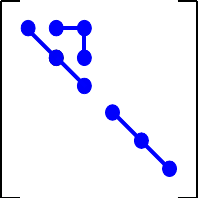}
  \end{center}
\end{figure}

which is exactly the form of the elasticity tensor for a cubic system.

% If your text is very short you might need to uncomment the following line to avoid
% layout problems with the figures and tables.
%\newpage

%%%%%%%%%%%%%%%% MAIN TEXT FIGURES %%%%%%%%%%%%%%%

%%%%%%%%%%%%%%%% MAIN TEXT TABLES %%%%%%%%%%%%%%%

%%%%%%%%%%%%%%%% REFERENCES %%%%%%%%%%%%%%%

\clearpage % Clear all remaining figures and tables then start a new page

% The list of references goes after the main text and before the acknowledgements
% When preparing an initial submission, we recommend you use BibTeX, like this:
%
\bibliography{Multivariant} % for a file named science_template.bib

% After the paper has completed peer review and been revised ready for acceptance,
% you should comment out the lines above and copy-paste the contents of your .bbl
% file here instead. This will help ensure that our conversion software works correctly.
% Remember to re-run BibTeX first - check the timestamp!
%
% Example of the first three entries copy-pasted from science_template.bbl:
%
%\begin{thebibliography}{1}
%
%\bibitem{example}
%A.~N. {Author}, An example reference. \emph{Journal of Improbable Research}
%  \textbf{1}, 67 (2020).
%
%\bibitem{example2}
%F.~M. {Surname}, S.~{Author}, A second example. \emph{Interesting Research
%  Letters} \textbf{32}, 897 (2019).
%
%\bibitem{example_preprint}
%P.~{One}, P.~{Two}, P.~{Three}, {An unpublished preprint}. \emph{preprint}
%  (2021), arXiv:2101.12345.
%
%\end{thebibliography}

%%%%%%%%%%%%%%%% ACKNOWLEDGEMENTS %%%%%%%%%%%%%%%

\section*{Acknowledgments}
H.A. thanks C. Barreteau for fruitful discussions. 

\section*{Funding}
This work was supported by ANR YOSEMITE: ANR-22-CE08- 0033-01.

%\paragraph*{Author contributions:}
%G.B. performed the STEM experiments, 
%H.A. performed the MC simulations for the Cu-Au system,
%J.N. participated to discussion and revision, 
%C.M. performed the MC simulations for the Co-Pt system,
%R.G. calculated the stress maps,
%J.C. optimized the SMA potential for Cu-Au,
%A.M. calculated the high resolution STEM images,
%D.A. analyzed sample composition by energy dispersive X-ray analysis,
%N.O.P. contributed to the analysis of the results,
%G.W. prepared the sample by epitaxy and
%C.R. written the paper.
%\paragraph*{Competing interests:}
%``There are no competing interests to declare.''
%\paragraph*{Data and materials availability:}
%All data are available in the main text or the supplementary materials.

%

%%%%%%%%%%%%%%%% SUPPLEMENT LIST %%%%%%%%%%%%%%%

% List the contents of your Supplementary Materials, including the numbers of any
% supplementary figures, tables, external data files etc. and any references that are
% cited only in the supplement. In this example, refs. 7-8 are cited only in the supplement.
% Fill out your numbers accordingly and delete any lines that aren't applicable.
\end{document}